# Polarization and wavelength agnostic nanophotonic beam splitter

David González-Andrade,[1,*] Christian Lafforgue,[2,3] Elena Durán-Valdeiglesias,[2] Xavier Le Roux,[2] Mathias Berciano,[2] Eric Cassan,[2] Delphine Marris-Morini,[2] Aitor V. Velasco,[1] Pavel Cheben,[4] Laurent Vivien,[2] and Carlos Alonso-Ramos[2]

[1]*Instituto de Óptica Daza de Valdés, Consejo Superior de Investigaciones Científicas (CSIC), Madrid 28006, Spain*
[2]*Centre de Nanosciences et de Nanotechnologies, CNRS, Université Paris-Sud, Université Paris-Saclay, C2N, Orsay 91405, France*
[3]*École Normale Supérieure Paris-Saclay, Université Paris-Saclay, Cachan 94230, France*
[4]*National Research Council Canada, 1200 Montreal Road, Ottawa, Ontario K1A0R5, Canada*
*Corresponding author: david.gonzalez@csic.es*

**High-performance optical beam splitters are of fundamental importance for the development of advanced silicon photonics integrated circuits. However, due to the high refractive index contrast of the silicon-on-insulator platform, state of the art Si splitters are hampered by trade-offs in bandwidth, polarization dependence and sensitivity to fabrication errors. Here, we present a new strategy that exploits modal engineering in slotted waveguides to overcome these limitations, enabling ultra-wideband polarization-insensitive optical power splitters, with relaxed fabrication tolerances. The proposed splitter relies on a single-mode slot waveguide which is transformed into two strip waveguides by a symmetric taper, yielding equal power splitting. Based on this concept, we experimentally demonstrate -3±0.5 dB polarization-independent transmission in an unprecedented 390 nm bandwidth (1260 - 1650 nm), even in the presence of waveguide width deviations as large as ±25 nm.**

The silicon-on-insulator (SOI) platform is becoming established as an enabling technology for next generation photonic circuits and a wide range of applications including telecom and datacom [1-3], radio-over fiber [4,5], bio-sensing [6,7], LIDAR [8] or absorption spectroscopy [9,10], to name a few. Such applications would benefit from the low cost and large volume fabrication of the existing CMOS facilities, as well as from the high density of integration enabled by the high refractive index contrast of the platform. Furthermore, the high modal confinement of Si wire waveguides provides strong light-matter interactions with a great potential for opto-electronic, sensing or non-linear optical devices [11]. However, the index contrast and modal confinement of the SOI platform also impose important challenges for the realization of high-performance SOI circuits, including strong sensitivity to small geometrical deviations, strong modal dispersion, and large birefringence between transversal electric (TE) and transversal magnetic (TM) modes. Hence, SOI circuits typically operate with a single polarization state, within a limited bandwidth and with tight fabrication tolerances. High-performance SOI fiber-chip interfaces have been demonstrated that yield wideband and high-efficiency coupling with negligible polarization dependence and relaxed fabrication tolerances [12-14], but this performance enhancement is still sought after in other essential SOI components. Beam splitters are particularly sensitive to the effects related to high index contrast and tight mode confinement in Si-wires, and would greatly benefit from achieving ultra-wideband dual-polarization operation.

Over the past years, several beam splitters have been proposed with different advantages and limitations. Directional couplers (DCs) allow simple tuning of the splitting ratio, but suffer from narrow bandwidth, low tolerances to fabrication errors and strong polarization dependence [15,16]. Constraints in fabrication tolerances have been alleviated by engineering the excitation of odd and even modes in shallow-etched [17] and fully-etched DCs [18-20], but only single-polarization operation or dual-polarization operation over a limited bandwidth have been demonstrated. Symmetric Y-junctions [21-23] provide a broad bandwidth and reduced polarization dependence, but exhibit comparatively high insertion loss due to the abrupt discontinuity at the junction. Multimode interference couplers (MMIs) [24] exploit the self-imaging (Talbot) effect to achieve low losses with improved fabrication tolerances, but provide a limited bandwidth and polarization-dependent behavior. Sub-wavelength engineering has been applied to increase MMI operational bandwidth [25,26], but only in single-polarization operation. Other alternatives include inverse tapers [27], star couplers [28], sub-wavelength engineered directional couplers [29-31], Y-branch variations [32-34], and adiabatic couplers [35], all of them restricted either in terms of limited operational bandwidth or demanding fabrication requirements.

In this Letter, we propose and experimentally demonstrate a new beam splitter concept based on modal-engineered slotted waveguides, enabling ultra broadband and polarization independent operation with relaxed fabrication tolerances. Our experimental results demonstrate a near-ideal transmission of -3±0.5 dB in an unprecedented bandwidth of 390 nm for both TE and TM polarizations.

The proposed beam splitter, schematically shown in Fig. 1, relies on a single-mode slot waveguide which equally splits the injected power into two output strip waveguides, in a wavelength- and polarization-agnostic fashion. Our device comprises three main sections: strip-to-slot mode converter (section I), slot waveguide (section II) and slot-to-strip splitting transition (section III). Single-mode operation of the slot waveguide is of fundamental importance to our device, as it mitigates any wavelength-dependent beating between different waveguide modes, which is the main mechanism limiting the bandwidth of DCs and MMIs. Adiabatic strip-to-slot mode converters ensure low-loss excitation of the fundamental (TE or TM) slot-waveguide mode [36,37], while the symmetry of the slot-to-strip transition ensures equal power splitting between the two output waveguides, independently of the polarization and the wavelength. Furthermore, this transition is robust against common under- or over-etching errors, as the latter do not break the structural symmetry, thus maintaining the power splitting ratio. Finally, as the slot waveguide already includes a central gap, our transition circumvents any abrupt index discontinuity, which is the major loss source in conventional Y-junctions.

In our design 220-nm-thick and 450-nm-wide ($W_I$) silicon-wires were used as the input and output strip waveguides. A buried silicon dioxide layer (BOX) and polymethyl methacrylate (PMMA) upper cladding were assumed, with refractive indexes $n_{Si} = 3.476$, $n_{SiO_2} = 1.444$ and $n_{PMMA} = 1.481$, respectively, at the central operating wavelength of 1.55 µm. First, we design the central slot waveguide in section II for single mode operation with no propagating higher order modes [38,39]. The slot waveguide comprises two narrow silicon waveguides, hereafter rails, separated by a low refractive index gap. We choose a rail width of $W_R = 150$ nm and a slot width of $G_S = 100$ nm, ensuring single-mode behavior in the wavelength range between 1200 nm and 1700 nm (see Fig. 2a). The strip-to-slot transition in the coupler (section I) is realized in two steps. In the first region of length $L_A$, a narrow rail progressively approaches the strip waveguide, while in the second region of length $L_B$, the strip waveguide and the narrow rail are combined into a slot waveguide. The rail at the beginning of the transition has a width of $W_J = 100$ nm and is separated from the strip waveguide by a gap of $G_A = 850$ nm, to preclude light coupling. After the first region of section I, the rail width is increased to $W_R = 150$ nm and the gap is reduced to $G_B = 250$ nm. As shown in Fig. 2b, low loss transition is achieved for both TE and TM modes by choosing a sufficiently long taper. Nevertheless, the slot length in section II is set to $L_S = 60$ µm in order to radiate away any residual power which is not coupled to the fundamental TE or TM slot waveguide modes. Finally, the slot-to-strip splitting transition (section III) is implemented by two symmetric trapezoidal tapers that increase the rail width from $W_R = 150$ nm to $W_I = 450$ nm and the slot width from $G_S = 100$ nm to $G_T = 350$ nm. A transition length of $L_T = 20$ µm was chosen for adiabatic behavior. Note that in the fabricated devices, the final separation between both output strip waveguides, $G_T$, is further increased to 1.75 µm by means of s-bends.

The performance and robustness against fabrication errors of the optimized device are studied by 3D EME simulations [40]. Effects of material dispersion are included in the analysis. As main figures of merit, we considered the excess loss (EL), defined as the amount of power relative to the input power that is not transferred to any output, and the imbalance (IB), defined as the power difference between the two output ports relative to the input power. Both figures of merit were calculated for the complete splitter (sections I, II and III) considering nominal design and ±25 nm deviations in waveguide width. Figure 3 shows the simulation results. Excess loss below 0.5 dB and imbalance below 0.3 dB in a 1200 nm - 1700 nm wavelength range are predicted for both the nominal and biased designs, confirming device resilience to fabrication errors. It should be noticed that the nominal design is a compromise between low insertion losses and a flat imbalance within the entire simulated bandwidth.

The beam splitter was fabricated in SOI wafers with a 220-nm-thick silicon and a 2 µm BOX layer. Electron beam lithography (Nanobeam NB-4 system, 80 kV) was used to define the splitter pattern. A dry etching process with an inductively coupled plasma etcher (SF6 gas) was employed to transfer the pattern to the silicon layer. Finally, 1-µm-thick PMMA was deposited as the upper cladding layer. Figure 4 shows scanning electron microscope (SEM) images of the fabricated beam splitter (before PMMA deposition).

Devices were characterized using three tunable lasers, covering sequentially the 1260 nm to 1650 nm wavelength range. Light was injected into the chip via 3-µm-wide waveguide edge couplers, using a lensed polarization-maintaining fiber, after polarization selection with a rotating quarter-waveplate and a linear polarizer. Light coming out of the chip was directed through a microscope objective and a linear polarizer to a photodetector. The splitter was characterized both using a Match-Zehnder interferometer (MZI) and cascaded stages. In the MZI configuration, two identical splitters were connected in back-to-back configuration, with an arm unbalance of 40 μm. Figure 5 shows the measured transmittance spectrum of the MZI interferometer for both TE and TM polarizations, compared to a reference strip waveguide. The device shows measured extinction ratio (ER) higher than 20 dB and 25 dB over a 390 nm wavelength range (1.26 – 1.65 μm) for TE and TM polarizations, respectively. The imbalance was calculated from measured extinction ratio [31,41], yielding better than ±0.42 dB for TE polarization and ±0.24 dB for TM polarization for the same wavelength range. Five stages of beam splitters were cascaded to carry out an accurate measurement of the transmission spectrum of the device. We included both the nominal design and biased structures with waveguide width variations of ±25 nm, allowing us determine tolerances to fabrication errors. One output of each splitter stage was collected at chip facet, whereas the second output was directed to the next splitter stage, enabling transmission measurement through linear regression fitting of the five resulting signals. As shown in Fig. 6, the slot-based splitters exhibit a deviation of less than ±0.5 dB with respect to the ideal -3 dB transmission for both polarizations within 1260 nm – 1650 nm wavelength range. The same robust performance is demonstrated even when introducing waveguide width variations up to ±25 nm.

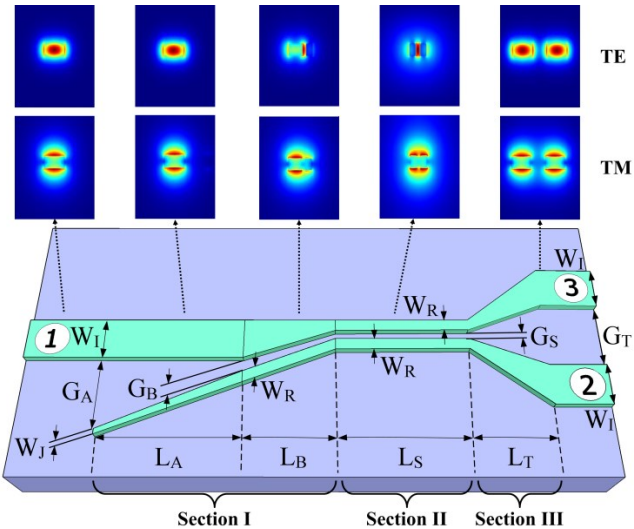

Fig. 1. Device schematic and main design parameters of the broadband polarization-independent beam splitter. Electric field mode profiles at different sections are shown for both TE and TM polarizations.

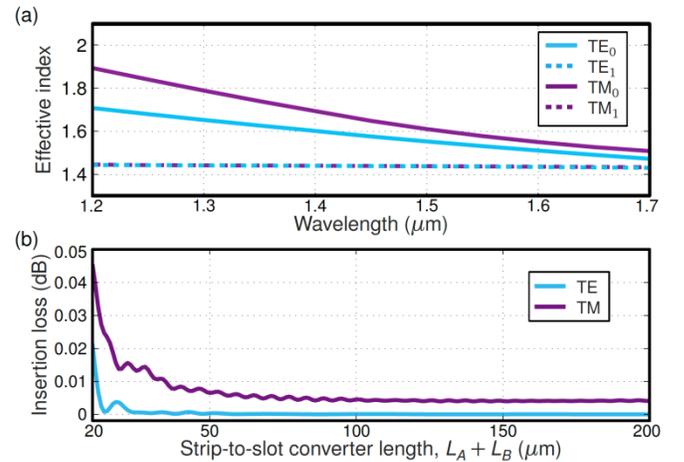

Fig. 2. (a) Effective index as a function of the wavelength for the fundamental and first-order TE and TM slot waveguide modes. Slot waveguide with $W_R = 150$ nm and $G_S = 100$ nm (see Fig. 1). Higher order modes with effective index near ~1.45 are below the cut-off condition. (b) Insertion loss as a function of the overall strip-to-slot converter length (Fig. 1 section I) for the fundamental TE and TM modes. The lengths $L_A$ and $L_B$ are varied maintaining a 2:1 ratio.

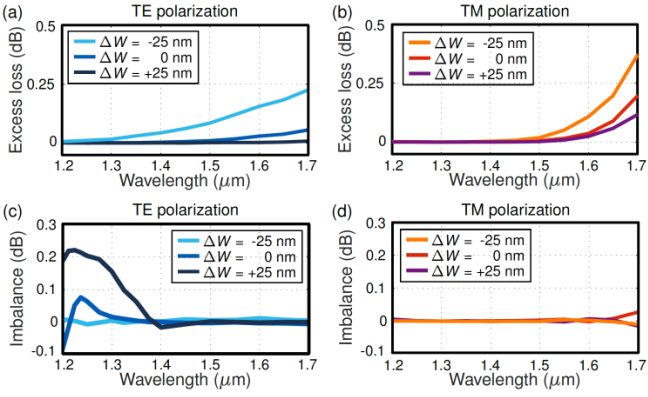

Fig. 3. Excess loss for TE (a) and TM (b) polarizations as a function of wavelength for the nominal design. Imbalance is also shown for TE (c) and TM (d) polarizations. We considered the nominal design ($\Delta W = 0$ nm) and deviations in all waveguide widths of $\Delta W = \pm 25$ nm.

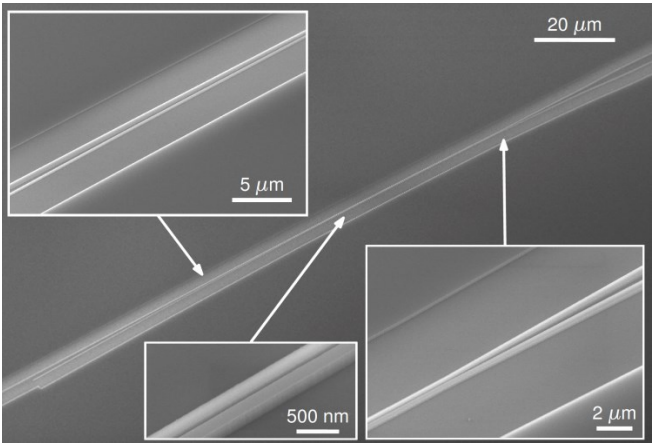

Fig. 4. Scanning electron microscope images of the fabricated splitter. Insets (left to right): strip-to-slot mode converter (section I), central slot waveguide (section II) and slot-to-strip output taper (section III).

In conclusion, we propose and experimentally demonstrate an ultra-broadband and polarization-independent optical beam splitter based on a single-mode slot waveguide with symmetric slot-to-strip transition. The single-mode operation of the slotted section prevents wavelength-dependent mode beating which fundamentally limits the bandwidth of conventional DC and MMIs, while the symmetric geometry ensures equal power split for both TE and TM polarizations even for comparatively large fabrication errors. The splitter yields a near ideal transmission of -3±0.5 dB for both polarizations in an unprecedented wavelength bandwidth of 390 nm, covering the O, E, S, C and L telecommunication bands. This outstanding performance is maintained in presence of fabrication deviations as large as ±25 nm, therefore relaxing fabrication constraints. We believe that the proposed optical beam splitter has an excellent potential for next generation of photonic integrated circuits as a key building block enabling dual-polarization and ultra-broadband silicon photonics devices.

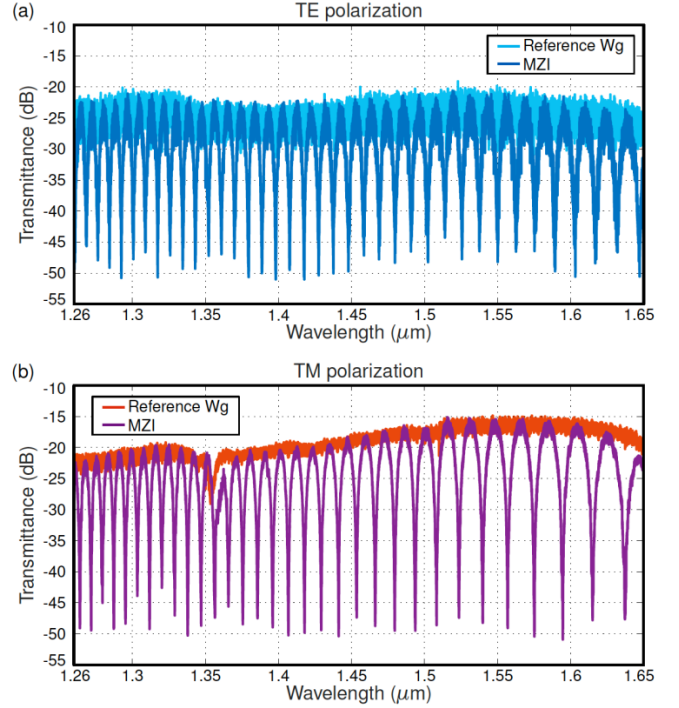

Fig. 5. Measured spectra of a Mach-Zehnder interferometer with back-to-back beam splitters, compared to a reference strip waveguide for TE (a) and TM (b) polarizations. All spectra are normalized to calibrate out setup loss.

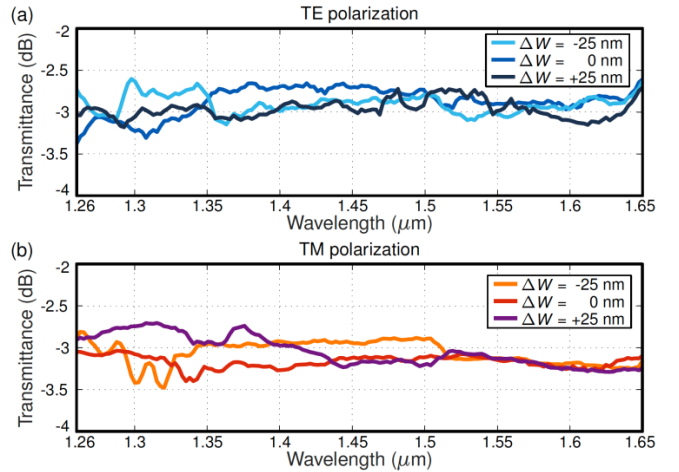

Fig. 6. Transmittance of splitters for TE (a) and TM (b) polarizations, measured through linear regression of the response of five cascaded stages.

**Funding.** Spanish Ministry of Science, Innovation and Universities (TEC2015-71127-C2-1-R with FPI scholarship BES-2016-077798, TEC2016-80718-R, IJCI-2016-30484); Community of Madrid (S2013/MIT-2790); EMPIR program (JRP-i22 14IND13 Photind), co-financed by the participating countries and the European Union's Horizon 2020 research and innovation program; Horizon 2020 research and innovation program (Marie Sklodowska-Curie 734331). H2020 ERC POPSTAR (647342). Agence National pour la Recherche, Project MIRSPEC (ANR-17-CE09-0041).

**Acknowledgements.** The sample fabrication has been performed at the Plateforme de Micro-Nano-Technologie/C2N, which is partially funded by the "Conseil Général de l'Éssonne". This work was partly supported by the French RENATECH network.